\documentclass[%
 reprint,
 superscriptaddress,
 amsmath,amssymb,
 aps,
 floatfix,
]{revtex4-2}
\usepackage{amsmath}
\usepackage{graphicx}
\usepackage[utf8]{inputenc}
\usepackage{dcolumn}
\usepackage{bm}
\usepackage[mathlines]{lineno}

\begin{document}

\preprint{APS/123-QED}

\title{Grain volume distribution alters the critical phenomena in complex granular systems}

\author{Teng Man}
\email{manteng0520@zjut.edu.cn}
\affiliation{
 College of Civil Engineering, Zhejiang University of Technology, Hangzhou, Zhejiang 310023, China
}%

\author{Yimin Lu}
\affiliation{
 Department of Civil, Environmental, and Construction Engineering, Texas Tech University, Lubbock, Texas 79409, United States
}

\author{Zhongrong Wang}
\affiliation{
 School of Engineering, Royal Melbourne Institute of Technology, Victoria, 3001, Australia
}%
 
\author{Herbert Huppert}%
\affiliation{%
 Institute of Theoretical Geophysics, King's College, University of Cambridge, King's Parade, Cambridge CB2 1ST, United Kingdom
}%

\author{Alessio Zaccone}
\affiliation{%
 Department of Physics “A. Pontremoli,” University of Milan, via Celoria 16, 20133 Milan, Italy
}


\author{Honglei Sun}
\email{sunhonglei@zju.edu.cn}
\affiliation{
 College of Civil Engineering, Zhejiang University of Technology, Hangzhou, Zhejiang 310023, China
}%


\date{\today}

\begin{abstract}
The grain size distribution (GSD) plays an important role in the mechanical properties of amorphous disordered systems and complex granular materials. Varying GSD causes segregation issues and alters critical behaviors. This work used the discrete element method (DEM) to investigate the rheological and critical behaviors of sheared granular flows with various GSDs. The results show that, while a unified rheological relation can be obtained, a characteristic length scale, which is associated with the contact probability and can be obtained from any GSD, is embedded within such a polydisperse disordered system. We further acquire a correlation function between critical solid fractions and dimensionless grain volume distributions. This work elucidates the effect of particle volumes on the rheology and micromechanics of dry granular systems and provides further insights in better incorporating the influence of other particle properties into a unified framework, which is helpful and critical for the corresponding engineering and geophysical problems.
\end{abstract}

\maketitle

Granular materials and disordered systems are commonly encountered in natural systems and industrial processes, such as landslides, pyroclastic flows, fresh concrete, and pharmaceutical and chemical particulates \cite{man2023mathematical,man2025jfm,lu2021flow,lu2024shear}. In recent decades, extensive investigations have been conducted to describe the rheology\cite{midi2004,Ge2023unifying}, compaction \cite{fiscina2010compaction,man2024materials}, and jamming transition \cite{song2008,bi2011jamming} of granular materials. The proposal of the inertial number $I_c$ and the viscous number $I_v$ suggests that the dynamics of granular flows is governed by the combination of three time scales: $t_i = d_p/\sqrt{\sigma_n/\rho_p}$, $t_v = \eta_f/\sigma_n$, and $t_M = 1/\dot{\gamma}$, where $t_i$, $t_v$, and $t_M$ are inertial, viscous, and macroscopic time scales, $d_p$ and $\rho_p$ are the particle diameter and density, $\sigma_n$ is the pressure, and $\eta_f$ denotes the fluid viscosity if the system is submerged or fully saturated.

However, the current rheological framework cannot fully capture the influence of either particle shapes or size distributions. The grain size distribution (GSD) plays an important role in the rheology of granular systems, but previous studies often provide seemingly contradictory opinions, where some studies suggested that the existence of small particles initiates a lubrication effect that helps the system increase its mobility \cite{ding2024grl}, while others believed that a wider GSD induces a denser packing fraction that increases the shear strength of granular systems and makes the system easier to transition to a jammed state \cite{Desmond2014PRE}. 

\citet{hill2011} studied the rheology of bidispersed granular flows using the discrete element model (DEM) \cite{yohannes2010,hill2011} and suggested that the pressure of a granular system is determined not only by its size distribution, but also by the effective free volume per particle, which was adopted from the jammed packing of monosized particles. They stated that the dimensionless pressure, $(1/I_c)^2$, scales as a function of the coordination number, $Z_c$. \citet{Polana2025monodisperse} studied the rheological behavior of polydispersed granular systems, which concluded that such systems still behave similarly to monodispersed systems if the right length scale is chosen, but the solid fraction can be quite different due to polydispersity. Recently, \citet{ding2024grl} also studied the behavior of granular systems with fractal distributions, and found that the shear strength of confined flow decreases at large fractal dimensions, which can be further linked to help decipher the mechanisms within the highly localized basal zones and the high mobility of general geological avalanches.


\begin{figure*}
	\centering
    	\includegraphics[scale=0.45]{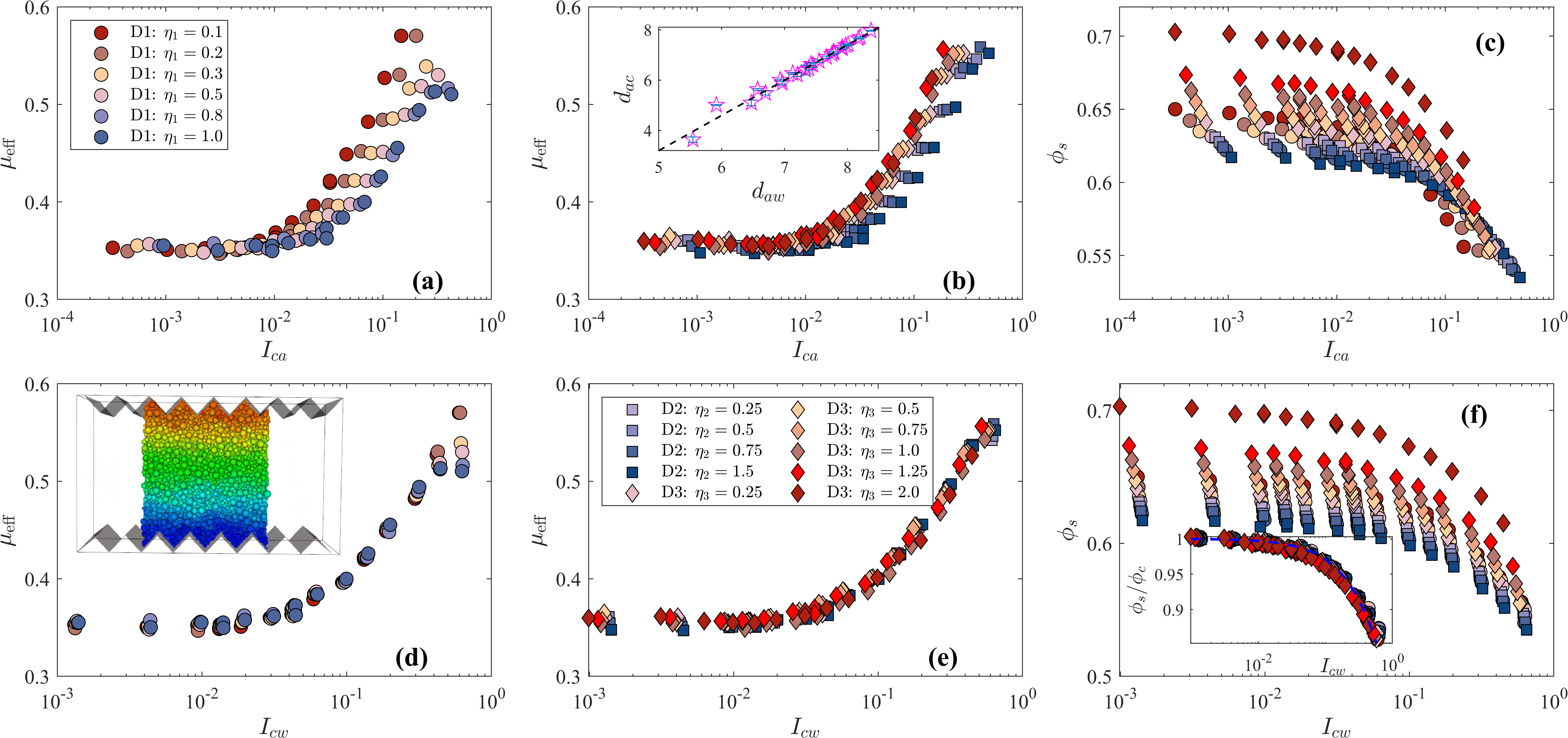}
    	\caption{\textbf{Rheological behaviors}: (a-c) The $\mu_{\mathrm{eff}}\sim I_{ca}$ and $\phi_s\sim I_{ca}$ relationships plotted for systems with size distribution D1, D2, and D3, where $I_{ca}$ is the conventional inertial number that uses system averaged particle diameter as the characteristic length. The inset of (b) shows the relationship between two length scales, $d_{ac}$ and $d_{aw}$, where $d_{ac}$ is the average particle diameter calculated based on contact number of each particle and $d_{aw}$ is the average particle diameter calculated based on volume of each particle. (e-f) The $\mu_{\mathrm{eff}}\sim I_{cw}$ and $\phi_s\sim I_{cw}$ relationships plotted for systems with size distribution D1, D2, and D3, where $I_{cw}$ is the new dimensionless number calculated with $d_{aw}$. Inset of (d) presents a simulation snapshop, and the inset of (f) is the relationship between $\phi_s/\phi_c$ and the proposed new inertial number $I_{cw}$ (see Equation~\ref{eq:Icw}.}
	\label{fig:1}
\end{figure*}

In this Letter, we focus on disordered physical systems with various continuous GSDs, where sheared granular materials with three different cumulative distribution functions (CDF) are examined using DEM. We show that, unlike the length scale proposed in \citet{Polana2025monodisperse}, a simply defined ``average" particle size, which has a clear physical root in the contact probability, already unifies the rheological behavior of dry granular systems with a wide range of different particle size distributions. Further analyses are devoted to better understand the critical phenomena when a granular system with different GSDs and frictional properties is approaching the ``jamming" transition.


\noindent\textbf{Methodology and simulation setup} In this work, the classical DEM is used to simulate sheared granular systems, where the particles are spherical and follow a linear spring-dashpot contact model. The normal contact force $F_{ij}^n$ is calculated as a Hookean contact with energy damping. The magnitude of the tangential contact force $|k_t\delta_{ij}^t|$ cannot exceed its friction limit, $\mu_p|F_{ij}^n|$, where $\mu_p$ is the particle frictional coefficient and for most cases we set $\mu_p=0.3$. The system has no rolling resistance. The granular system is located between two sawteeth plates and is sheared with constant pressure and shear rate. For different simulations, we vary the pressure $\sigma$ from 50 to 1000 Pa and the shear rate $\dot{\gamma}$ from 0.01 to 1.0 s$^{-1}$, and measure the mean value of the key variables when the sheared system is already in steady state, shown in Appendix \ref{sec:append1}, where its solid fraction $\phi_s$, pressure $\sigma$, shear stress $\tau$, and velocity profile become stable. We can obtain the effective frictional coefficient $\mu_{\mathrm{eff}} = \tau/\sigma$, the effective viscosity $\eta_{\mathrm{eff}} = \tau/\dot{\gamma}$, and the average inertial number $I_c$ from each simulation for further investigation. The plot of a shear cell is shown in the inset of FIG. \ref{fig:1}(d). The particle size distributions used in this work, i.e., Fuller distribution and two power-law distributions, are elaborated in detail in Appendix \ref{sec:append2}.

\noindent\textbf{Rheological behaviors} According to the classical $\mu(I)$ rheology of dry granular systems, both $\mu_{\mathrm{eff}}=\tau/\sigma$ and $\phi_s$ should scale with the inertial number $I_{ca} = \dot{\gamma}d_p/\sqrt{\sigma/\rho_p}$, where $d_p$ is the particle diameter and also a characteristic length scale. \citet{Ge2023unifying} proposed that the length scale plays an important role in determining the rheological behavior of granular systems, and we further argue that this length scale should be adjusted to reflect changes in particle size distributions. FIGs. \ref{fig:1}(a-b) demonstrate the $\mu_{\mathrm{eff}}\sim I_{ca}$ relationship for systems with GSDs of D1, D2, and D3, which are clarified in Appendix \ref{sec:append2}, whereas FIG. \ref{fig:1}(c) shows the $\phi_s\sim I_{ca}$ relationship, where $I_{ca}$ represents the conventional inertial number that uses the average particle diameter of the system as the characteristic length.

FIGs. \ref{fig:1}(a-c) shows that, even though the $\mu_{\mathrm{eff}}\sim I_{ca}$ and $\phi_s\sim I_{ca}$ relationships collapse for systems with any specific GSD, changing the particle gradation leads to deviations in both $\mu_{\mathrm{eff}}$ and $\phi_s$. FIGs. \ref{fig:1}(a) and (b) show that systems with different particle gradations result in different transitional $I_{ca}$, which denotes a transition from quasi-static to inertial flow. The change in this transitional $I_{ca}$ implies that the characteristic length of the system may also change due to changes of GSDs. \citet{hill2011} reported that the average particle size should be calculated based on each particle volume. Thus, we define the characteristic length of the system as $d_{aw} = [\sum_{i=1}^{N_p}( d_{pi}V_{pi})] / (\sum_{j=1}^{N_p}V_{pj})$ and particles with different volumes are assumed to have different weight in calculating this ``average" particle size. Thus, a new inertial number can be defined as
\begin{equation}
    \begin{split}
        I_{cw} = \dot{\gamma}d_{aw}(\rho_p/\sigma)^{1/2}\ .
    \end{split}
    \label{eq:Icw}
\end{equation} 

After adopting the new inertial number, as shown in FIGs. \ref{fig:1}(d,e), the relationships between $\mu_{\mathrm{eff}}$ and $I_{cw}$ collapse onto a single master curve, indicating that the characteristic length, $d_{aw}$, reflects some physical and topological nature of the sheared granular systems, and choosing an appropriate length scale helps unify the frictional rheology of dry granular systems with any size distribution. However, we admit that $I_{cw}$ fails to unify the $\phi_s\sim I_{cw}$ relationship, although FIG. \ref{fig:1}(f) looks much better than FIG. \ref{fig:1}(c). For systems with different GSDs, their transitional $I_{cw}$ seems to be unified, and the only difference lies in the $\phi_s-$axis, i.e., moving the $\phi_s\sim I_{cw}$ relationships upward or downward can also unify this relation. For a series of simulations with the same GSD, the relationship between $\phi_s$ and $I_{cw}$ can be written as $\phi_s = \phi_c/(1+\alpha_{\phi}I_{cw})$, where $\alpha_{\phi} \approx 0.32$. Thus, if we change the vertical coordinate into $\phi_s/\phi_c$, where $\phi_c$ is the critical solid fraction for each GSD, we can obtain a master curve of $\phi_s/\phi_c = 1/(1+\alpha_{\phi}I_{cw})$, shown as the blue dashed curve in the inset of FIG. \ref{fig:1}(f).

\begin{figure}
	\centering
    	\includegraphics[scale=0.40]{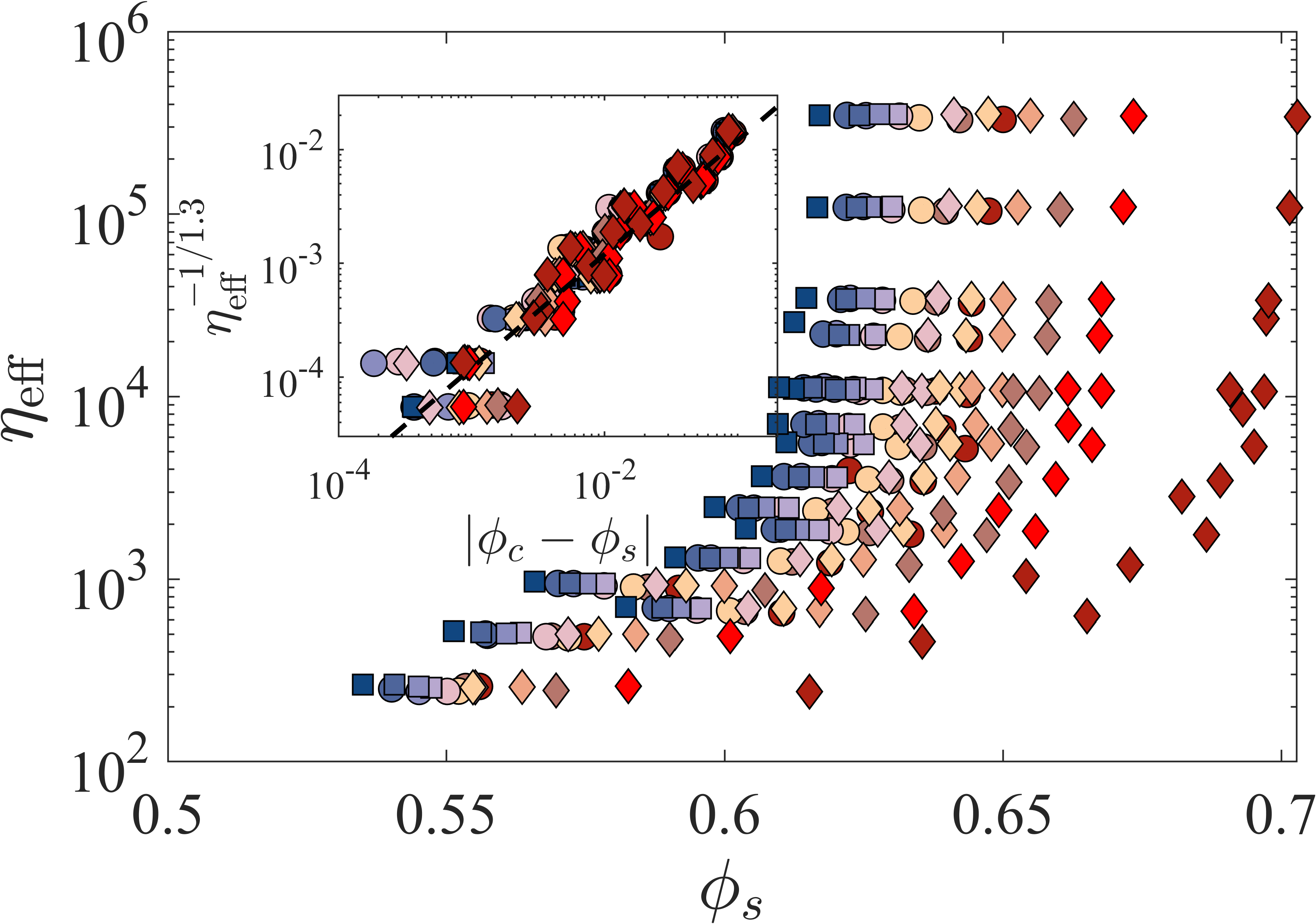}
    	\caption{Relationship between the effective viscosity of granular systems, $\eta_{\mathrm{eff}}=\tau/\dot{\gamma}$, and the solid fraction, $\phi_s$. In the inset of this figure, we change the horizontal and vertical axes into $|\phi_s - \phi_c|$ and ${\eta_{\mathrm{eff}}}^{-1/1.3}$, respectively. The dashed line represents a linear relationship.}
	\label{fig:2}
\end{figure}

\noindent\textbf{Critical length scale and critical solid fraction} Our choices of characteristic length and critical solid fraction have deep physical and geometrical reasons, even though they seem to be deliberately fitted. On one hand, $d_{aw}$ can be related to the coordination number of each polydispersed granular system. The characteristic length should be linked to particle volumes because they are directly related to the number of contacts. Intuitively, a particle with larger size has a greater chance to contact with more particles, indicating that such a particle plays a more important role in bearing strong force networks and determining the macroscopic behavior of the whole system. To validate this, we calculated another length scale, $d_{ac}$, based on contact numbers of each particle, where $d_{ac} = [\sum_i(d_{pi}N_{ci})]/(\sum_j N_{cj})$ and $N_{ci}$ is the contact number of particle $i$, and plotted its relationship with $d_{aw}$ in the inset of FIG. \ref{fig:1}(b). This shows that $d_{aw}$ and $d_{ac}$ are linearly related, which verifies that $d_{aw}$, which can be easily obtained from GSD, has a deep root in contact statistics.

$\phi_c$ varies with the GSD and frictional properties of particles. It is clearly neither the jamming solid fraction, $\phi_{J}$ \cite{bi2011jamming}, nor the random close packing fraction, $\phi_{RCP}$ \cite{Zaccone2025JAP}. However, \citet{Zaccone2025JAP} and his follow-up research on the random close packing of the polydisperse granular system \cite{Zaccone2025PT} encourage us to argue that $\phi_c$ is a function of $\phi_{RCP}$ and $\mu_p$. Since we kept $\mu_p$ constant, $\phi_c$ is predominantly controlled by $\phi_{RCP}$, which should be related to the GSD of a granular system. Thus, we could obtain a functional relationship between $\phi_c$ and the GSD. 

\begin{figure*}
	\centering
    	\includegraphics[scale=0.45]{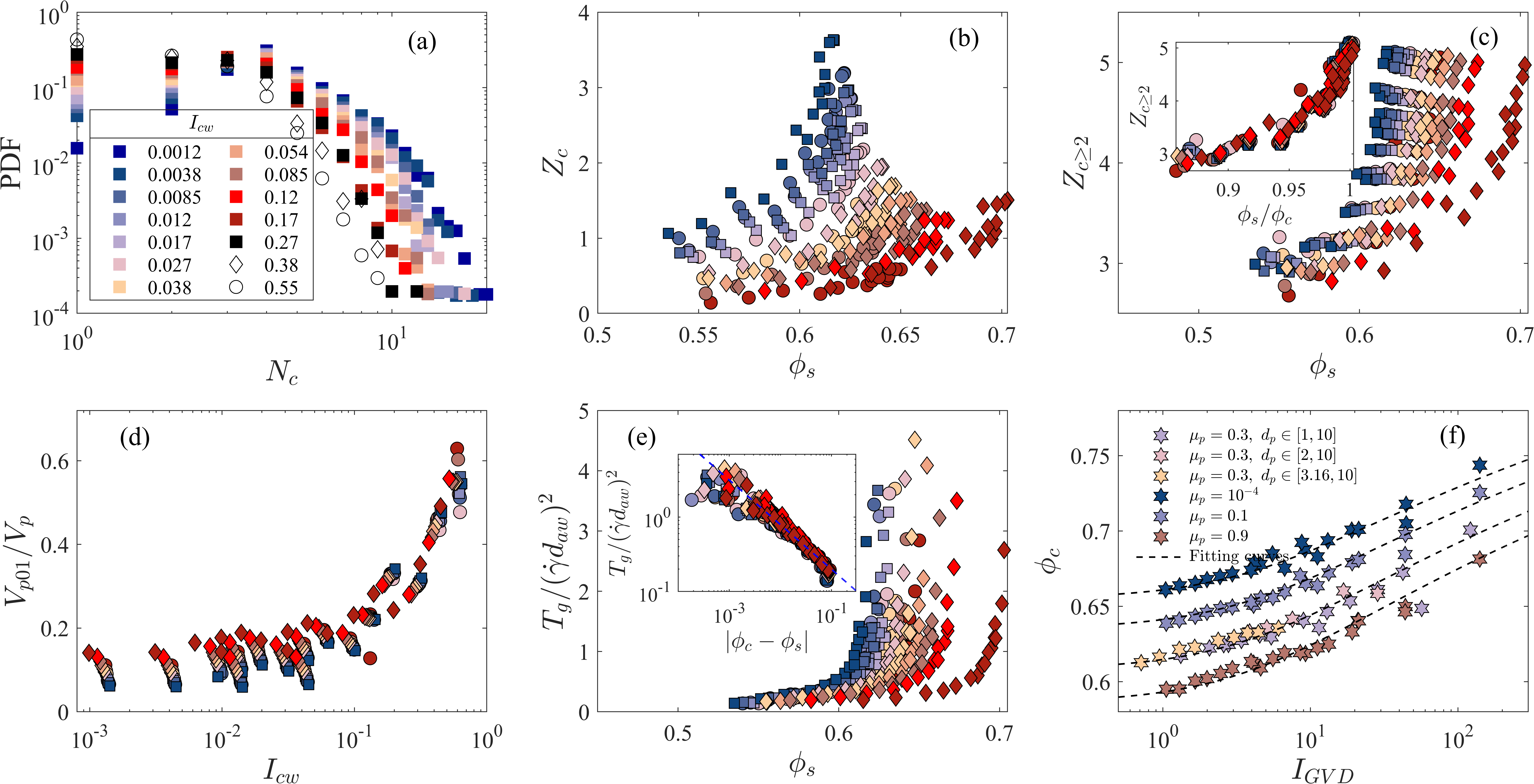}
    	\caption{(a) Histogram of the number of contact while changing $I_{cw}$. This figure only plots data from the system of D3 and $\eta_s = 1.0$. (b) The $Z_c\sim \phi_s$ relationship for systems with different GSDs, and $Z_c$ is the average coordination number. (c) The volume percentage of particles with only 0 or 1 contact is plotted against $I_{cw}$. (d) Relationship between the dimensionless granular temperature, $T_g/(\dot{\gamma}d_{aw})^2$, and $\phi_s$ (inset: $T_g/(\dot{\gamma}d_{aw})^2\sim|\phi_c - \phi_s|$). (e) The relationship between $\phi_c$ and the dimensionless index of the grain volume distribution, $I_{GVD}$.}
	\label{fig:3}
\end{figure*}

We note that $\phi_c$ for each GSD should not be obtained from fitting of the $\phi_s\sim I_{cw}$ relationship, since the functional form of the $\phi_s\sim I_{cw}$ relationship is based on phenomena, not physics. Fortunately, \citet{Zaccone2025JAP} derived the relationship between effective viscosity $\eta_{\mathrm{eff}}=\tau/\dot{\gamma}$ and $\phi_s$ from first principles, which yields $\eta_{\mathrm{eff}}\sim |\phi_c - \phi_s|^{-1.3}$. Thus, if we plot ${\eta_{\mathrm{eff}}}^{-1/1.3}$ against $|\phi_c - \phi_s|$, we should be able to obtain a linear relationship. In FIG. \ref{fig:2}, we plot $\eta_{\mathrm{eff}}$ against $\phi_s$ for all simulations and find that $\eta_{\mathrm{eff}}$ always diverges at relatively high solid fractions (which are different for systems with different GSDs). The solid fraction, where a granular system is approaching infinite viscosity, indicates a critical solid fraction, $\phi_c$. For each GSD, we can obtain a specific $\phi_c$, and ${\eta_{\mathrm{eff}}}^{-1/1.3}$ and $|\phi_c - \phi_s|$ are linearly related, as shown in the inset of FIG. \ref{fig:2}.

Our above analyses indicate that the length scale $d_{aw}$, which implies contact statistics, helps to unify the $\mu_{\mathrm{eff}}\sim I_{cw}$ relationship but leaves a problem of $\phi_c$ for us to solve. This implies that, for systems with different GSDs but exactly the same $d_{aw}$, their stress-strain rate relationships behave similarly, but their solid fractions differ from each other. Since stresses in granular materials are strongly related to force networks (especially the strong force network), granular systems with different GSDs may have different amounts of rattler particles, which do not alter the strong force network, but occupy different amounts of volumes, resulting in different solid fractions.

We define particles with zero contact or only one contact as rattlers, while particles with more than one contact as non-rattlers. While calculating the contact number of each particle, $N_c$, and the coordination number, $Z_c\equiv\langle N_c\rangle$, of each simulation, we can also obtain the volumetric percentage of rattlers, $V_{p01}/V_{p}$, and the coordination number of non-rattlers, $Z_{c\geq 2}$. FIG. \ref{fig:3}(a) plots the histograms for a set of simulations with GSD of D3 and $\eta_3 = 1.0$, where the legend provides the information of $I_{cw}$. Increasing $I_{cw}$ leads to a decrease in the number of particles with large contact numbers and increases the chances of finding particles with only one contact, indicating that increasing the inertial number results in more binary collisions. With such a wide range of particle sizes, a sheared granular system naturally has many ``small" particles, which do not play a role in the force network. FIG. \ref{fig:3}(b) indicates that, for some cases, $Z_c$ is even below 1.0, but for each set of simulations with the same GSD, the data collapse onto a $Z_c\sim\phi_s$ master curve. However, it is difficult to figure out how changing the GSD alters $Z_c$. 

Neglecting particles with only zero and one contact number and changing the vertical axis into the average coordination number of particles with more than one contact, $Z_{c\geq 2}$, in FIG. \ref{fig:3}(c), the relationship $Z_{c\geq 2}\sim \phi_s$ shows power-law features with systematic changes introduced by changing the GSD. The inset of FIG. \ref{fig:3}(c) shows that dividing $\phi_s$ by $\phi_c$ helps to get a unified $Z_{c\geq2}\sim\phi_s/\phi_c$ relationship, which is similar to the $\eta_{\mathrm{eff}}\sim\phi_s/\phi_c$ relationship. We argue that it is mostly particles with $N_c\leq 2$ that determine the effective viscosity of the system. All of these imply that changing the GSD leads to a different percentage of particles that participate in bearing forces in granular systems. However, FIG. \ref{fig:3}(d) contradicts our expectation and shows that changing the GSD does not change the volume percentage of particles much, as $N_c\leq 1$. 

We further plot the relationship between the normalized granular temperature $T_g/(\dot{\gamma}d_{aw})^2$ and $\phi_s$ in FIG. \ref{fig:3}(e), where $T_g \equiv \langle (\vec{u} - \langle \vec{u} \rangle)^2 \rangle$ and $\vec{u}$ is the particle velocity. Again, the critical solid fraction $\phi_c$, which is heavily influenced by the GSD, plays a key role in the scaling of $T_g/(\dot{\gamma}d_{aw})^2\sim\phi_s$, and $T_g/(\dot{\gamma}d_{aw})^2$ scales with $\sim |\phi_c-\phi_s|^{-0.6}$. We argue that $\phi_c$, as a critical solid fraction, marks the transition from a system with finite viscosity to that with infinite viscosity. Physically, it has similar characteristics as random close packing fractions $\phi_{RCP}$, which are affected by both $\mu_p$ and GSD. 

\citet{Zaccone2025PT} and \citet{Anzivino2023} suggested that $\phi_{RCP}$ can be expressed as a function of the standard deviation of GSD, while \citet{Desmond2014PRE} argued that they could draw a unified relation between $\phi_{RCP}$ and a combination of skewness and polydispersity of GSD with the data obtained from simulations. However, we find that both methods do not work for our system. We adopted the idea of \citet{Desmond2014PRE} and calculated both the skewness $S_v$ and the polydispersity $\delta_v$ of the normalized grain volume distribution (GVD), instead of the GSD, so that
\begin{subequations}
    \begin{align}
        \tilde{V}_p^i = \left[V_p^i - \min(V_p)\right]/\left[ \max(V_p)-\min(V_p) \right]\ , \\
        S_v = \langle \Delta\tilde{V}_p^3 \rangle/\left(\langle \Delta\tilde{V}_p^2 \rangle\right)^{3/2}\ ,\ \delta_v = \sqrt{\langle \Delta\tilde{V}_p^2 \rangle}/\langle \tilde{V}_p \rangle\ ,
    \end{align}
\end{subequations}
where $\tilde{V}_p$ is the normalized particle volume, $\Delta\tilde{V}_p = \tilde{V}_p - \langle \tilde{V}_p\rangle$, $\langle \tilde{V}_p^n \rangle = \int \tilde{V}_p^n\cdot P(\tilde{V}_p) d\tilde{V}_p $, and $\langle \Delta\tilde{V}_p^{(n)} \rangle = \int \Delta\tilde{V}_p^n\cdot P(\tilde{V}_p) d\tilde{V}_p$. To obtain a universal relationship, we varied both the size ranges ([1, 10] cm, [2, 10] cm, and [3.16, 10] cm) and the particle frictional coefficient ($\mu_p = $0.0001, 0.1, 0.3 and 0.9). With a combined dimensionless index of the GVD, $I_{GVD}=\delta_v + S_v\delta_v^2$, we obtained an empirical relationship between $\phi_c$ and $I_{GVD}$ so that
\begin{equation} \label{eq:phic}
    \begin{split}
        \phi_c =& 1-(1-\phi_{c0})/\left[1 + \beta\ln{(1 + I_{GVD}/I_{g0})} \right]\ ,
    \end{split}
\end{equation}
where $\beta\approx 0.087$ and $I_{g0} = 4.5$ are fitting constants, $\phi_{c0}$ is the base $\phi_c$ and varies with $\mu_p$, and the functional form of Eq. \ref{eq:phic} is inspired by the time evolution of the solid fraction of granular system under tapping compactions in \citet{knight1995density}. FIG. \ref{fig:3}(f) suggested that we only need to adjust $\phi_{c0}$ to get a reasonable $\phi_c\sim I_{GVD}$ relationship for systems with different $\mu_p$, and Eq. \ref{eq:phic}, although being empirical, introduces a logarithmic dependence that ensures the eventual ``flattening" of $\phi_c$ with increasing $I_{GVD}$. Ultimately, $\phi_c$ should flatten out to a ``maximum" packing fraction that is lower than or equal to 1 when $I_{GVD}\rightarrow +\infty$, as also suggested in previous publications \cite{Desmond2014PRE,Anzivino2023,Hoy2025}. In Appendix \ref{sec:append3}, we plot the $\phi_{RCP}\sim I_{GVD}$ relationship for data extracted from previous publications to show that Eq. \ref{eq:phic} also works for predicting the random close packing fractions.

This Letter focuses on the rheology of granular systems with different grain size distributions and identifies a characteristic length scale that can be associated with the contact probability. We find that varying the GSD leads to the change in criticality behavior, i.e. the critical solid fraction $\phi_c$ that marks the transition from a finite $\eta_{\mathrm{eff}}$ to an infinite one. The analyses of $Z_c$ and $T_g$ verify the importance of determining $\phi_c$ based on the GVD, and we have introduced a dimensionless index, $I_{GVD}$, to successfully acquire a unified solution for granular systems with different particle gradations, which has the potential to be adopted to other athermal disordered systems with various polydispersities. 

\begin{acknowledgments}
\noindent\textit{Acknowledgments-} We wish to acknowledge the financial support of Grant No. 12202367 from NSFC and the start-up grant from Zhejiang University of Technology. T.M. thanks Prof. K.M. Hill, Prof. J.-L. Le, Dr. Z. Ge, and Prof. S.A. Galindo-Torres for helpful discussions. The simulations in this work were conducted on an open-access multiphysics simulation library: MECHSYS, which can be acquired from Github (\url{https://github.com/Axtal/mechsys.git})
\end{acknowledgments}

\appendix
\label{appendix}

\begin{figure}
	\centering
    	\includegraphics[scale=0.42]{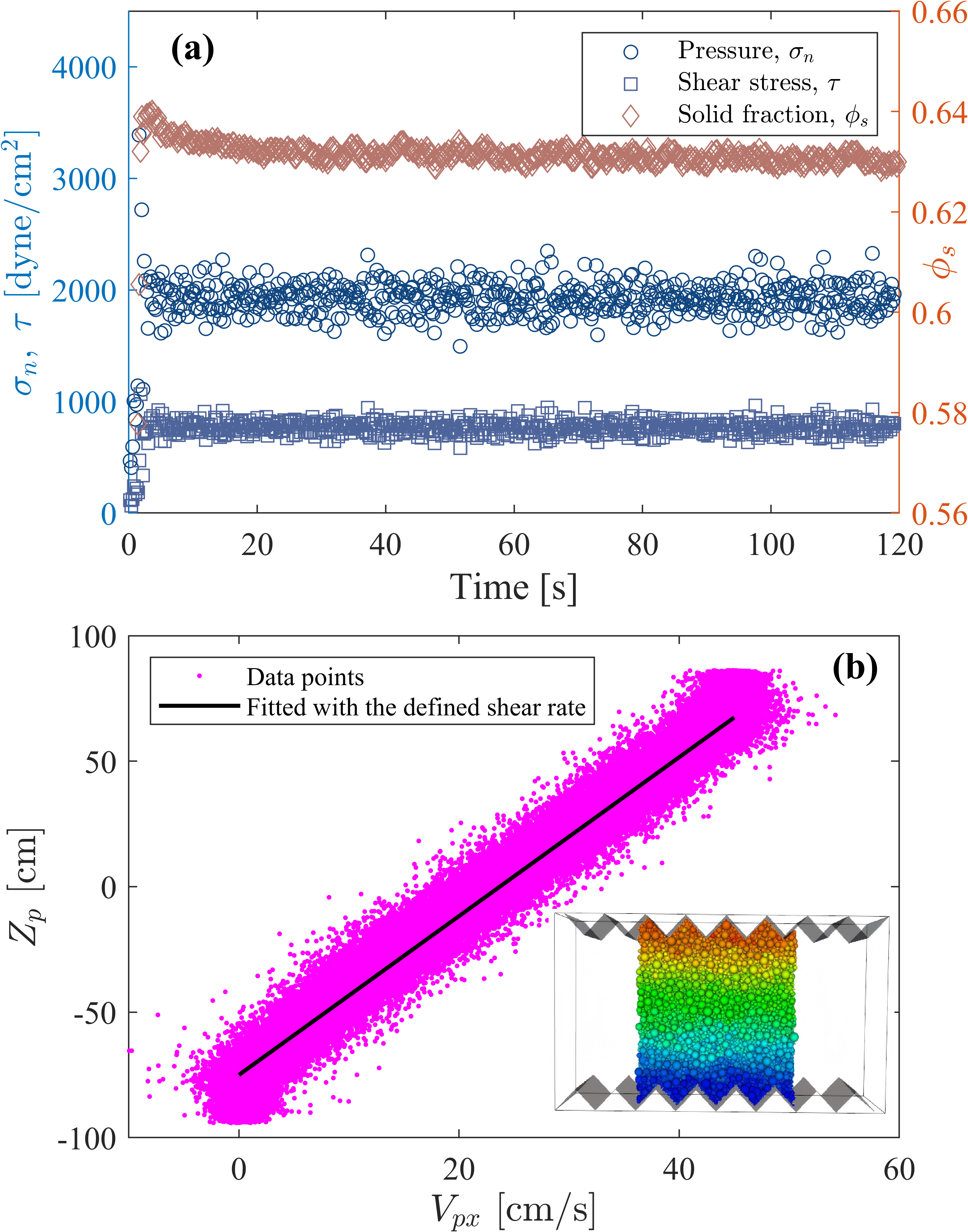}
    	\caption{(a) The time evolution of the shear stress, $\tau$, pressure, $\sigma_n$, and the solid fraction, $\phi_s$. (b) the profile of particle velocities in the $x-$direction, $V_{px}$. The inset of (b) shows the configuration of a simple shear simulation of the granular system.}
	\label{appfig:1}
\end{figure}

\begin{figure*}
	\centering
    	\includegraphics[scale=0.45]{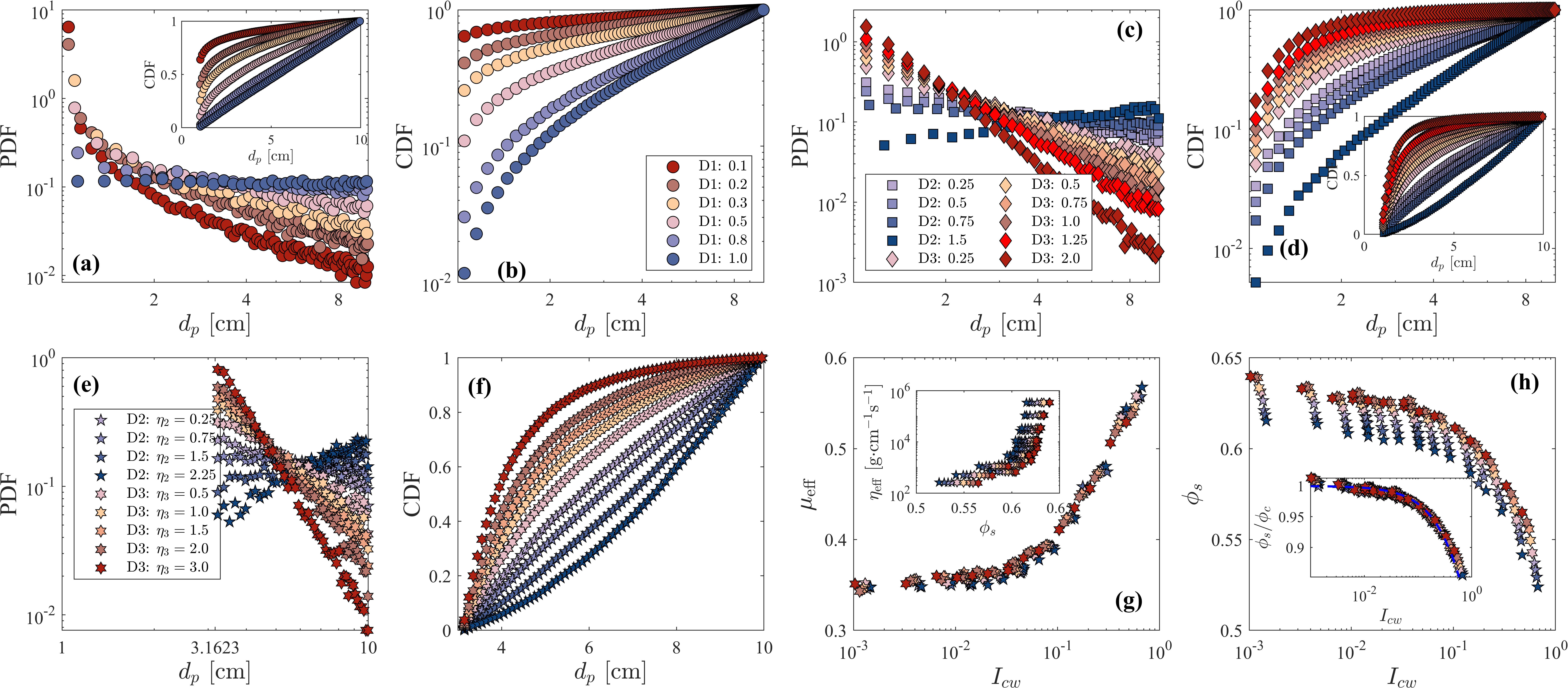}
    	\caption{(a-d) show both the PDF and CDF of different GSDs for particle size ranging from 1 to 10 cm, while (c) and (f) show the PDF and CDF for systems with $d_p\in [3.16, 10]$ cm. In Figs. (g,h), we plot the rheological properties of systems with $d_p\in [3.16, 10]$ cm.}
	\label{appfig:2}
\end{figure*}

\section{Simulations and stress calculations}
\label{sec:append1}

The discrete element method (DEM) solves the momentum balance equation for both translational and rotational motions. The governing equations of DEM can be written in the form of the Newton's second law, where the calculation of contact forces is of vital importance. In each simulation, the governing equations are integrated using a Verlet method \cite{Torres2010}, which is commonly used in molecular dynamics.  Particles are spherical and follow a linear spring-dashpot contact model, where the normal contact force $F_{ij}^n$ is calculated as a Hookean contact with energy damping, so that $F_{ij}^n = -k_n\delta_{ij}^n-c_n\dot{\delta}_{ij}^n$, where $k_n$ is the contact stiffness, $c_n$ is the damping coefficient, and $\delta_{ij}^n$ is the contact overlap. $F_{ij}^n$ has a viscous term for energy dissipation, where $c_n$ is calculated based on the coefficient of restitution, $e_n$, so that
\begin{equation}
    \begin{split}
        c_n = 2 \xi_n \sqrt{m_{\mathrm{eff}}k_n}\ ,\xi_n = -\ln{e_n} / \sqrt{\pi^2+(\ln{e_n})^2}\ ,
    \end{split}
\end{equation}
where $m_{\mathrm{eff}} = (1/m_i+1/m_j)^{-1}$ is the reduced mass and we set $e_n=0.5$. The magnitude of the tangential contact force $|k_t\delta_{ij}^t|$ cannot exceed its frictional limit, $\mu_p|F_{ij}^n|$, where $\mu_p$ is the friction coefficient of particles. The granular system is located within and sheared by two sawteeth plates at constant pressure and shear rate, as shown in the inset of FIG. \ref{appfig:1}(b). At each timestep, the sheared granular system reaches a steady state when $\phi_s$, $\tau$, and $\sigma_n$ become stable. Meanwhile, instead of measuring stresses from the upper and lower plates, we calculated the average stress tensor for each output timestep. The average stress tensor $\sigma_{ij}$ can be obtained by
\begin{equation}
    \begin{split}
        \sigma_{ij} = (1/V)\sum_c^{N_{ct}}{f_c^i l_c^j}\ ,
    \end{split}
\end{equation}
where $V$ is the volume of the granular system, $f_c^i$ is the $i$th component of a contact force vector (a summation of normal and tangential contact forces of the $c$th contact), and $l_c^j$ is the $j$th component of a contact branch vector that links centroids of two contacting particles. FIG. \ref{appfig:1}(a) presents typical time-lapse evolutions of $\phi_s$, $\tau$, and $\sigma_n$, showing a clear steady state for determining the representative average values. Since we utilized sawteeth plates as upper and lower boundaries, the velocity profile across height is always linear [shown in FIG. \ref{appfig:1}(b)], indicating a constant shear rate.

\section{Particle size distributions}
\label{sec:append2}

In this letter, we use two different types of particle size distributions, i.e., Fuller distribution (D1) and power-law distributions (D2 and D3). The cumulative distribution function (CDF) of D1 can be written as
\begin{equation} \label{eq:D1}
    \begin{split}
        C_1(d_p) = \left[ (d_p - d_{\min})/(d_{\max}-d_{\min}) \right]^{\eta_1}\ .
    \end{split}
\end{equation}
Then, a random particle diameter that follows this CDF can be obtained by imposing a uniformly distributed random number within $[0, 1]$, $\mathrm{Rand}(0,1)$, on the left-hand side of Eq. \ref{eq:D1} so that
\begin{equation}
    \begin{split}
        d_p = d_{\min} + [\mathrm{Rand}(0,1)]^{1/\eta_1}\cdot(d_{\max}-d_{\min}).
    \end{split}
\end{equation}
Considering the possible slopes of the PDF of power-law distributions, we take two different functional forms:
\begin{subequations}
    \begin{align}
        &C_2(d_p) = (d_p/d_{\max})^{\eta_2}\ , \\
        &d_p = d_{\max}\cdot[\mathrm{Rand}(0,1)]^{1/\eta_2}\ ,\\
        &C_3(d_p) = 1-(d_p/d_{\min})^{-\eta_3}\ ,\\ 
        &d_p = d_{\min}\cdot[1-\mathrm{Rand}(0,1)]^{-1/\eta_3}\ .
    \end{align}
\end{subequations}
where we impose the $d_p\in [d_{\min}, d_{\max}]$ condition every time we generate a random particle diameter. Figures \ref{appfig:2}(a-f) show both the PDF and CDF of systems with different GSDs, and FIGs. \ref{appfig:2}(g, h) show the rheological properties of granular systems with a different size range, which indicate that changing the particle size range would not change the rheological behavior, especially when we consider $I_{cw}$ and eliminate the effect of $\phi_c$.

\begin{figure}
	\centering
    	\includegraphics[scale=0.42]{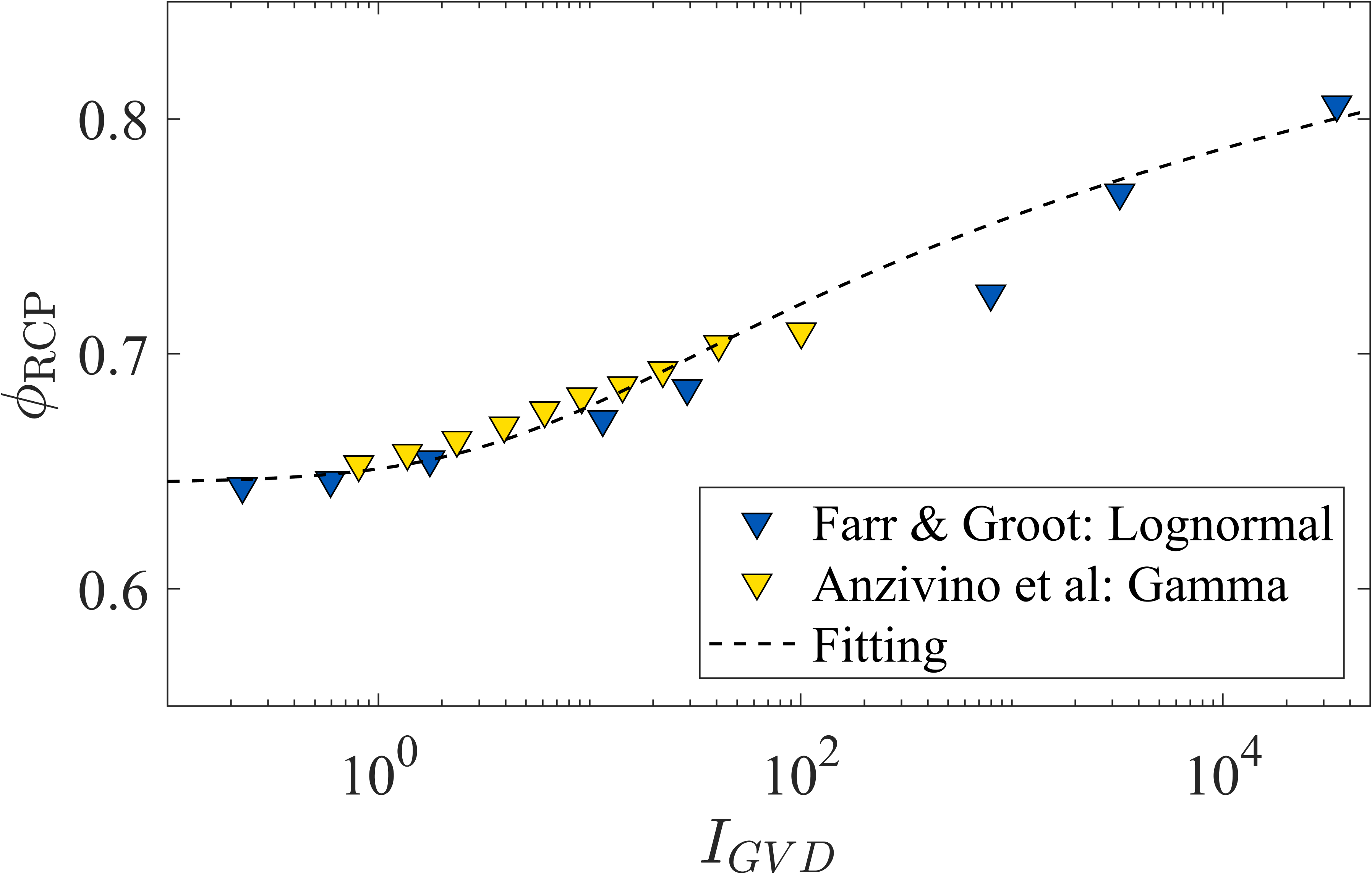}
    	\caption{Data extracted from published papers, where the particle size in \citet{Farr2009} follows a lognormal distribution and that in \citet{Anzivino2023} follows a Gamma distribution. The published data are fitted with Eq. \ref{eq:phic}.}
	\label{appfig:3}
\end{figure}

\section{$\phi_{RCP}$ extracted from previous publications}
\label{sec:append3}

We mentioned that previous research has focused on predicting the random close packing solid fraction, $\phi_{RCP}$, which is closely related to $\phi_c$ in our present work. However, we make clear that $\phi_{RCP}$ is different from $\phi_c$. The studies of $\phi_{RCP}$ are mostly focused on frictionless systems. Furthermore, DEM simulations, where particles are in loading conditions, allow for the existence of overlaps among contacting particles, which will result in relatively larger solid fractions. To validate the $\phi_c\sim I_{GVD}$ relationship, we turn to published works to also link $\phi_{RCP}$ to $I_{GVD}$. In FIG. \ref{appfig:3}, we plot the data extracted from both \citet{Farr2009} with lognormal size distribution and \citet{Anzivino2023} with $\Gamma$ distribution. We converted their particle size distributions into particle volume distributions and obtained the corresponding statistical parameters to calculate $I_{GVD}$. FIG. \ref{appfig:3} shows that our empirical relationship of Eq. \ref{eq:phic} also works for predicting $\phi_{RCP}$ for frictionless spheres with a single fitting parameter of $\phi_{c0}\approx 0.645$.


\bibliography{partgradation}

\end{document}